\documentclass[12pt]{iopart}
\usepackage{graphicx}
\usepackage{bm}
\usepackage{afterpage}

\newcommand{\bol}[1]{\boldsymbol #1}
\begin{document}

\title
[A short guide to topological terms in the effective theories of condensed matter]
{A short guide to topological terms in the effective theories of condensed matter}

\author{Akihiro Tanaka \& Shintaro Takayoshi}

\address
{
Computational Materials Science Unit, 
National Institute for Materials Science, 
Namiki 1-1, Tsukuba, Ibaraki, Japan, 305-0044
}
\ead{
TANAKA.Akihiro@nims.go.jp
}
\vspace{10pt}
\begin{indented}
\item[]July 2014
\end{indented}

\begin{abstract}

This article  
is meant as a gentle introduction to 
the  ``topological terms'' that 
often play a decisive role in effective theories describing 
topological quantum effects in  
condensed matter systems.  
We first take up several prominent examples,  
mainly from the area of quantum magnetism and superfluids/superconductors.  
We then briefly discuss how these ideas are now finding 
incarnations in the studies of symmetry-protected 
topological phases, which are in a sense the generalization of the 
concept of topological insulators to a wider range of materials, including 
magnets and cold atoms. 
\end{abstract}


\section{Introduction}

Topological quantum effects, whose significance to material functions  
is gaining wide recognition with the discovery of the topological insulator, 
have actually been (along with electron correlation effects) one of the 
core pillars of contemporary condensed mater physics since the 1980s. 
It thus seems appropriate to try to achieve an understanding of 
problems where topology plays a governing role 
more or less as a whole: such an attempt will hopefully help fit the ongoing topics 
taken up in this issue into a larger physical frame, while the older problems can be revisited in the light of the latest developments. 

The purpose of this article is to take a microscopic step toward that goal 
by taking a cursory glance at the role played by 
{\it topological terms}, which typically show up in the low-energy action of  
systems where topological quantum effects take place. The emphasis is  
on pedagogy, and no attempt is made on completeness or rigor. 
Our modest objective is to warm the interested nonspecialist 
reader to some of the basic features of topological terms 
through simple examples, preparing her/him to be exposed to the vast literature 
on this rapidly evolving subject.  
  
There are  monographs which 
cover similar grounds but from a broader perspective\cite{Auerbach,Fradkin,Nagaosa,Altland}, 
to which we refer the reader for an in-depth exposition. Altland and Simons\cite{Altland} in particular 
offers a readable digression on the defining mathematical structures 
of topological terms, on which will not detail here.  

\section{The two uses of path integrals in condensed matter}

The format we employ below is that of path integrals, which is used in  
a good many theoretical work on this subject. 
Aside from being the natural language for modern quantum field theory, which is 
where topological terms originates from, it proves to be particularly convenient  
when it comes to extracting the low energy properties of a many body system, 
especially those of a topological nature.    

There are mainly two ways in which path integrals come into play in our discussion. 
One is in evaluating the probability amplitude of a quantum mechanical process, 
i.e. a complex number 
whose squared modulus yields the probability. 
Suppose we are studying a field $\phi$ 
(generally a function of time $t$ and position $\bol{r}$) 
which, in our examples below, is typically an order parameter for some condensed matter 
(such as a superconductor or a ferromagnet).  
If we are interested in the 
probability amplitude associated with 
the realization of 
a certain spatial configuration $\phi_{\rm f}(\bol{r})$, 
we can formally write it down as the path integral (for most of the following we will set 
$\hbar=h/2\pi\equiv 1$)
\begin{eqnarray}
{\rm Amp}
&=&
\langle\phi_{\rm i}\vert e^{-i\int_{t_{\rm i}}^{t_{\rm f}} dt{\cal H}}\vert\phi_{\rm f}\rangle
\nonumber\\
&=&\int_{\phi_{\rm i}(\bol{r})\rightarrow \phi_{\rm f}(\bol{r})} {\cal D}\phi(t, \bol{r})e^{i{\cal S}[\phi(t, \bol{r})]}.
\label{path integral}
\end{eqnarray}
The first line is the textbook expression for this quantity which involves 
the time evolution kernel $e^{-it{\cal H}}$, where ${\cal H}$ is the 
Hamiltonian of the system (suitable generalizations are to be made when ${\cal H}$ at different times do not commute). In the path integral formula in the second line, 
 the integration stands for a summation over all possible ways in which 
$\phi$, which at an initial time took the configuration $\phi_{\rm i}(\bol{r})$, evolves 
at the final time into the configuration $\phi_{\rm f}(\bol{r})$.  Each version 
of these evolution histories is called a path, 
and is weighted by the Feynman weight $e^{i{\cal S}}$, where 
${\cal S}$ is the action. In what follows it will often prove to be convenient to 
convert from real time ($t$) to the imaginary time ($\tau\equiv it$) formalism. 
Although mathematically this can be a highly nontrivial procedure, 
the formal prescription for this conversion is actually rather simple and 
consists of two steps. First we make the substitution $t=-i\tau$ in eq.(\ref{path integral}). 
Following this, we {\it define} the imaginary time action ${\cal S_{\rm E}}$ using the original 
real time action ${\cal S}_{\rm M}$ action (the subscripts each stand for Euclidean and 
Minkowskian space-time) through the correspondence $e^{i{\cal S}_{\rm M}}\vert_{t=-i\tau} 
\equiv e^{-{\cal S}_{\rm E}}$. Although we will suppress the subscripts E and M below for 
the sake of notational brevity, 
we will specify which of the two frameworks we are working with. 
It can be shown\cite{Kogut} 
that by continuing to imaginary time 
and then 
evaluating the above path integral expression for the amplitude,which now takes the form 
$\int {\cal D}\phi(\tau, \bol{r}) e^{-{\cal S}[\phi(\tau, \bol{r})]}$, we can effectively project our theory  
onto the ground state. 
We will later take advantage of this fact to evaluate the ground state wavefunctional $\Psi[\phi(\bol{r})]$, i.e. the probability 
amplitude for the configuration $\phi(\bol{r})$ to be  realized in the ground state. 
 
 The second usage is to gain information  on 
the partition function ${\cal Z}={\rm Tr}e^{-\beta {\cal H}}$ which arises in quantum statistical mechanics. To convert to a path integral representation, one breaks up the Boltzmann weight 
into a product of  many exponential operators, 
$e^{-\beta{\cal H}}= (e^{-\Delta \tau{\cal H}})^{N}$, where $\Delta \tau \equiv \frac{\beta}{N}$. Since each $e^{-\Delta \tau{\cal H}}$ can be viewed as the generator of 
time evolution over a interval $\Delta \tau$ in imaginary time, the partition 
function can be written, on taking the limit $N\rightarrow\infty$, as the imaginary time path integral 
\begin{equation}
{\cal Z}=\int_{\phi_{\rm i}=\phi_{\rm f}} {\cal D}\phi(\tau, \bol{r})e^{-{\cal S}[\phi(\tau, \bol{r})]}. 
\end{equation}
Here we are summing over paths in which the field $\phi(\tau, \bol{r})$ 
obeys a periodic boundary condition in the imaginary-time direction, 
which reflects the trace operation that is involved in calculating the partition function. 

The definition of a topological term ${\cal S}_{\rm top}$ varies considerably among authors. 
For instance, one can adopt the view that they are the actions which are {\it metric-independent}, i.e. do not change form with the topology of space-time\cite{Altland}. 
Here we use the terminology in its broadest sense, and associate it with 
actions which have the 
following two features. (1)It is the portion of the 
action which arises {\it in addition to} the kinetic action  
coming directly from the Hamiltonian ${\cal H}$, 
i.e. the total action should generally have the form 
${\cal S}={\cal S}_{\rm kin}+{\cal S}_{\rm top}$, where (in imaginary time) 
 ${\cal S}_{\rm kin}=\int d\tau {\cal H}$.  
(2) The term ${\cal S}_{\rm top}$, when using  the imaginary time framework 
is {\it purely imaginary} and hence contributes a {\it phase factor} to the ``Boltzmann'' 
weight $e^{-{\cal S}}$. (For a simple single particle quantum mechanics example on this aspect see Auerbach\cite{Auerbach} or Altland-Simons\cite{Altland}.) 
The first feature implies that the topological term usually 
has no apparent classical counterpart, while the second feature is suggestive of 
nontrivial quantum interference effects which may arise when an action contains a 
topological term. Thus it is often an indication of interesting physics.  
It is also often the case, as justifies its name, 
that ${\cal S}_{\rm top}$ bears a clear topological 
significance, such as a proportionality to a topological winding number. 
We will encounter a number of such examples in the following. 

Deriving, or establishing the existence of a topological term in the 
low energy effective theory of a system starting from a given Hamiltonian can be 
a subtle and highly technical issue. (For example, whether or not a topological term 
known as the Hopf term should arise in the effective action of 2d Heisenberg antiferromagnets 
was briefly controversial\cite{Fradkin}.) Our plan therefore will be to examine without going into the derivation itself, illustrative examples where topological terms are known to exist, and explore how their presence will alter the low energy physics.

\section{Quantum interference}

Suppose that the total effective action of the system in consideration 
consists of a kinetic term plus a topological term, i.e.   
${\cal S}={\cal S}_{\rm kin}+{\cal S}_{\rm top}$. As mentioned in the previous section, 
the probability amplitude 
associated with the transition $\vert 1 \rangle \longrightarrow \vert 2 \rangle$ 
is, in path integral language, ${\rm Amp}=\int _{1\rightarrow 2}[{\rm path}]e^{i{\cal S}}$, 
where the integration symbol simply means that a summation over all paths 
for which the system is in state $\vert 1\rangle$ at initial time and state 
$\vert 2 \rangle$ at final time be taken.  
Important consequences can arise if this summation is dominated by paths 
which cost the system essentially the same kinetic energy, but differ in 
the value of the topological term. The amplitude then factorizes into the form 
\begin{equation}
{\rm Amp}\approx e^{i{\cal S}_{\rm kin}}(e^{i{\cal S}_{\rm top}[{\rm path A}]}+
e^{i{\cal S}_{\rm top}[{\rm path B}]}+e^{i{\cal S}_{\rm top}[{\rm path C}]}+\cdot\cdot\cdot).
\end{equation}  
Clearly the topological term can lead to a quantum interference effect: in particular, 
if the factors $e^{i{\cal S}_{\rm top}[{\rm path}]}$ sum up to zero (i.e. a destructive interference), the transition in question is forbidden, even though the 
transition amplitude for each path $e^{i{\cal S}}$ is nonzero. 
(This scenario applies for both 
the real time and the imaginary time path integral. )
We now take up in the following several examples where this happens. 

\subsection{Single spin: nanomagnet}
Consider a spin moment with spin quantum number $S$. 
Suppose that there is an dominant easy-plane and a sub-dominant easy-plane 
anisotropy; we take for concreteness the Hamiltonian ${\cal H}=-J_{1}S_{z}^{2}+J_{2}S_{y}^{2}$, 
with $J_{1}>J_{2}>0$. If we denote the direction of the spin vector by the unit vector $\bol{n}$ 
(the direction is only meaningful in a semi-classical sense since the spin components 
are non-commuting entities), the action is given by 
\begin{eqnarray}
{\cal S}&=&{\cal S}_{\rm B}+{\cal S}_{\rm kin}
\nonumber\\
&=& S\omega[\bol{n}(t)]+\int dt{\cal H}.
\end{eqnarray}
The topological term ${\cal S}_{\rm B}$ is related to the 
Berry phase which is induced by the evolution of the spin orientation.  
If the motion of $\bol{n}(t)$ is such that its orientation coincides at the 
beginning and the end of the time interval under consideration, 
this term can be expressed as in the 
second line, where $\omega[\bol{n}(t)]$is the solid angle, i.e. the area 
which the trajectory of $\bol{n}(t)$  
traces out on the surface of the unit sphere.  
In terms of the spherical coordinate $\bol{n}=(\sin\theta\cos\phi, 
\sin\theta\sin\phi, \cos\theta)$, this is explicitly written as 
\begin{equation}
{\cal S}_{\rm B}=S\int dt (1-\cos\theta(t))\partial_{t}\phi(t).
\end{equation}
\begin{figure}[h]
\begin{center}
\includegraphics[width=3.5in]{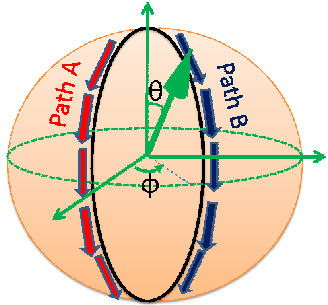}
\caption{
The two low-energy paths relevant to the evaluation of the transition amplitude 
from the up-spin state to the down-spin state. The difference of the Berry phase 
for the two paths is well defined and equals $S$ multiplied by $2\pi$, i.e. the surface area corresponding to half of the unit sphere. 
}
\label{fig:nanomagnet}
\end{center}
\end{figure}
Let us now consider the probability amplitude 
associated with the transition of $\bol{n}$ from the north to the south pole, 
i.e. between the two lowest energy states\cite{Loss,von Delft}. 
The path integral for this transition is dominated (see Fig. 1) by the two low energy paths connecting the poles, which are located along the longitudes $\phi=0$ (path A) and $\phi=\pi$ (path B). Though 
each path  alone does not trace out a closed curve, their {\it difference} does. Noting that 
the two paths correspond to the same kinetic energy, we expect 
the transition amplitude to take the form 
\begin{eqnarray}
{\rm Amp}&\approx& e^{i{\cal S}[{\rm path A}]}+e^{i{\cal S}[{\rm path B}]}
\nonumber\\
&=&e^{i{\cal S}_{\rm kin}}e^{i{\cal S}_{\rm B}[{\rm path A}]}
\left(
1+e^{i\left( {\cal S}_{\rm B}[{\rm path B}]-{\cal S}_{\rm B}[{\rm path A}]\right)}
\right)\nonumber\\
&=&
 e^{i{\cal S}_{\rm kin}}e^{i{\cal S}_{\rm B}[{\rm path A}]}
\left(
1+e^{i2\pi S}
\right),
\end{eqnarray}
where, in the third line, we have used the fact that the surface area corresponding to 
half of the sphere is $2\pi$. (A detailed evaluation can be found in 
the original references\cite{Loss,von Delft}.) 
Thus if $S$ is half of an odd integer ($S=1/2, 3/2, ...$), the probability amplitude vanishes (destructive interference), while if $S$ is integer-valued ($S=1, 2, ...$), 
the amplitude is enhanced (constructive interference).  This can have experimental consequences, for example for 
switching effects in single-molecule nanomagnets such as Mn${}_{12}$-acetate and Fe${}_8$.  
The two interference patterns discussed above can manifest themselves as 
unavoided or avoided level crossings between the up-spin and down-spin levels 
(the former case corresponds to the suppression of hybridization) 
as a function of an external magnetic field applied along the z-axis. 
(It has been pointed out however that realistic situations tend to obscure this effect\cite{MiyaNaga}.)   
It is also interesting that essentially 
the same mechanism has been proposed to control the macroscopic tunneling 
of magnetic flux in a fabricated superconducting island (it is possible to map the low energy 
sector onto an effective spin system), which may find applications 
to quantum information processes\cite{Friedman}.

\subsection{Superconductor and superfluid}
We turn to the effective low energy action of a superfluid or a superconductor;  
we will focus on the former as most of what follows can be generalized 
to the latter by simply coupling the system to an electromagnetic field. 
Assuming that the amplitude of the condensate's order parameter is well-developed, 
the low energy physics can be described solely in terms of the phase degree of freedom $\phi$, 
and the effective action typically takes the form\cite{Fisher Magnus} 
(with applications to the partition function in mind, we 
employ the imaginary time formalism)
\begin{eqnarray}
{\cal S}&=&{\cal S}_{\rm top}+{\cal S}_{\rm kin}\nonumber\\
&=&
i\int d\tau d\bol{r}\rho\partial_{\tau}\phi+\int d\tau d\bol{r}\frac{K}{2}
\left[(\partial_{\tau}\phi)^2 +  (\nabla\phi)^2 \right] .
\label{quantum XY action}
\end{eqnarray} 

\begin{figure}[h]
\begin{center}
\includegraphics[width=3.5in]{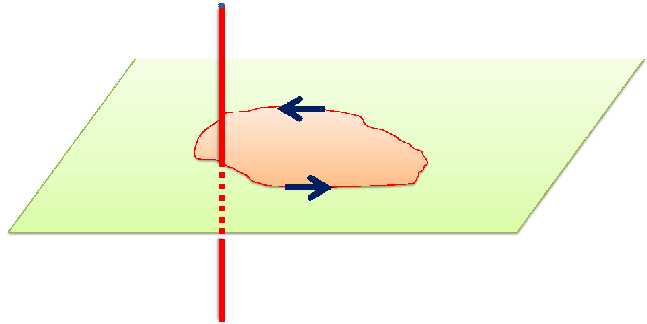}
\caption{
The trajectory of a vortex executing a round excursion 
through a superfluid (of superconductor) thin film. 
The Berry phase associated with this process, which is recorded by the 
topological term,  
is proportional to the number of particles within the area bounded by 
the closed curve. In analogy with the Aharonov-Bohm effect, this 
indicates that the vortex behaves like a charged particle in an external 
magnetic field. 
}
\label{fig:vortex1}
\end{center}
\end{figure}

The second term on the right hand side can be derived from a standard Ginzburg-Landau 
type action using a phase-only approximation.  
(The coefficient $K$, which describes the rigidity against phase fluctuations, is proportional 
to the square of the amplitude of the order-parameter. We also note that we have 
set for simplicity a coefficient with the  dimension of a velocity to unity.) 
The first term is the topological 
or Berry phase term, and resembles the ``$ip\partial_{\tau}q$''-term which 
appears in the (imaginary-time) 
Lagrangian of a single particle whose dynamics is described in terms of a 
pair of canonically conjugate variables $q$ (position) and $p$ (momentum). Recalling that 
the canonical conjugate of the phase $\phi$ 
is the particle density of the superfluid condensate, we interpret 
the coefficient $\rho$ as the superfluid density (or more precisely, the   
offset value that this term imposes on this physical quantity). To see how this term 
influences the physics of the superfluid, 
we will study the motion of a vortex moving about in the system 
-for simplicity we consider a 2d system, i.e. a superfluid thin film, with a constant 
value of $\rho$. Furthermore, 
we assume that the vortex returns after an excursion to its initial 
position (see Fig. 2), as is required from the periodicity in the imaginary time direction.  
It is easy to check that the action ${\cal S}_{\rm top}$ contributes 
the quantity $i2\pi q_{\rm v}$ each time the vortex goes around a (bosonic) particle in a superfluid 
condensate (or in superconductor language, a Cooper pair), 
where $q_{\rm v}\in{\bf Z}$ is the vorticity. Hence, if the total number of 
bosons which has been encircled by the vortex is 
$N_{\rm b}=\rho A$, where $A$ is the area bounded by the trajectory (note that in the 
2d case $\rho$ is the number of condensate particles per area, and $N_{\rm b}$ generally 
is {\it not} an integer), 
the net outcome from the topological term becomes 
${\cal S}_{\rm top}=i2\pi N_{\rm b}=i2\pi \rho A$. 
In other words, it contributes to the ``Boltzmann weight'' $e^{-{\cal S}}$ 
entering the path integral a phase factor of 
\begin{equation}
e^{-{\cal S}_{\rm top}}=e^{-i2\pi N_{\rm b}}=e^{-i2\pi \rho A}.
\label{vortex phase factor SF}
\end{equation} 
This may be viewed as an Aharonov-Bohm-like effect; vortices see the surrounding 
condensate particles as a sort of magnetic field (with an intensity proportional to $\rho$), 
as is clear from the  
phase accumulation that occurs when the vortex performs a round trip.

\begin{figure}[h]
\begin{center}
\includegraphics[width=3.5in]{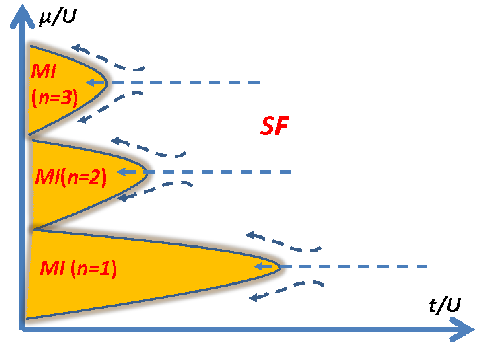}
\caption{
A schematic phase diagram of the boson Hubbard model. SF and MI stand for the 
superfluid and Mott insulator phases. The dotted arrows denote the constant $\rho$ 
contours. When the boson filling factor is an integer, i.e. $\rho\in {\bf Z}$, 
the system can undergo a SF to MI transition 
upon reducing the parameter $t/U$. For generic filling factors, the superfluid phase 
persists. This can be understood in terms of a destructive interference among 
vortex excitations due to the presence of the topological term. }
\label{fig:bosehubbard}
\end{center}
\end{figure}

The phase factor of eq.(\ref{vortex phase factor SF}) can lead to important consequences. Generally, when a condensation of vortex excitations 
in a superfluid condensate (point vortices in 2d, vortex loops in 3d, space-time vortices, or 
phase-slip events in 1d) occurs, it will destroy the superfluidity, and the system 
is expected to enter a new phase. While we are familiar with 
these types of phase transitions that occur at finite 
temperatures, they can also happen at zero temperature 
with the variation of some control parameter.    
Quantum phase transitions of this variety 
can be studied in detail using the boson Hubbard model, 
which recently attracts much attention due to its direct relevance to 
the physics of cold atoms. 
This model has the following lattice Hamiltonian\cite{Sachdev},  
\begin{equation}
{\cal H}=-t\sum_{\langle ij\rangle}b^{\dagger}_{i}b_{j}-\mu\sum_{i}n_{bi}+U\sum_{i}n_{bi}(n_{bi}-1).
\end{equation}
The summation in the first term is to be taken with respect to nearest neighbor sites, 
$b_{i}$ ($b^{\dagger}_{i}$) is a boson annihilation (creation) operator, $n_{bi}=b^{\dagger}_{i}b_{i}$ is the boson number 
operator, and $\mu$ and $U$ are each the chemical potential and the onsite Coulomb repulsion. Advances in cold atom experiments now enables researchers to access a wide 
region of the phase diagram of this model, which is depicted in Fig. 3. 
When $t\gg U$, the system is in a superfluid phase, for which the effective 
action of eq.(\ref{quantum XY action}) gives a suitable description. When $t/U$ is reduced, 
the tendency to form a superfluid weakens, and one naively expects to encounter a 
phase transition into a Mott insulator phase. However,  as already noted this transition proceeds by the condensation of vortices, which are each accompanied 
by the Berry phase factors 
of the form written in eq.(\ref{vortex phase factor SF}). For generic (irrational) values of 
$\rho$, which translates in the lattice model to generic boson filling factors, these will 
enter into the path integral as 
{\it random phase factors}. Although further analysis 
is necessary to work out the details\cite{Sachdev}, it is then reasonable to deduce 
 that configurations containing  vortices will cancel out and effectively drop out from the partition function altogether, 
meaning that the system cannot undergo a transition into the insulating phase. 
An exception occurs however when $\rho\in{\bf Z}$, i.e. at integer boson filling, since for this case the phase factor of eq.(\ref{vortex phase factor SF}) is unity. 
This expectation can be confirmed by a study of the boson Hubbard model. 
At integral filling, a transition into the Mott insulator phase occurs at the 
tip of the Mott insulator phase lobe, while for generic filling factors, the 
constant-$\rho$ contour avoids entering the insulator phase and 
escapes into the region between the lobes. (The case where the filling factor is a 
nonintegral rational number requires a careful treatment -the possibility that 
vortices with some higher vorticity will condense, giving rise to an exotic phase, 
has to be taken into account-  
and is beyond the scope of the present argument.) 
Although we have focused in the above on a 2d system, 
it is straightforward to show that similar reasonings apply to systems of other 
dimensionalities. 

\subsection{Magnetization plateau}
A closely related phenomenon, also driven by the phase interference between 
vortex Berry phases, occurs in antiferromagnets in an external magnetic field. 
When one probes the magnetization per site $m$ 
as a function of the external field $H$, 
there often appear plateau regions in the magnetization curve (see Fig. 4), contrary 
to classical analysis which predicts a monotonic increase. The crucial question is 
when and how the plateau forms.

\begin{figure}[ht]
\begin{center}
\includegraphics[width=3.5in]{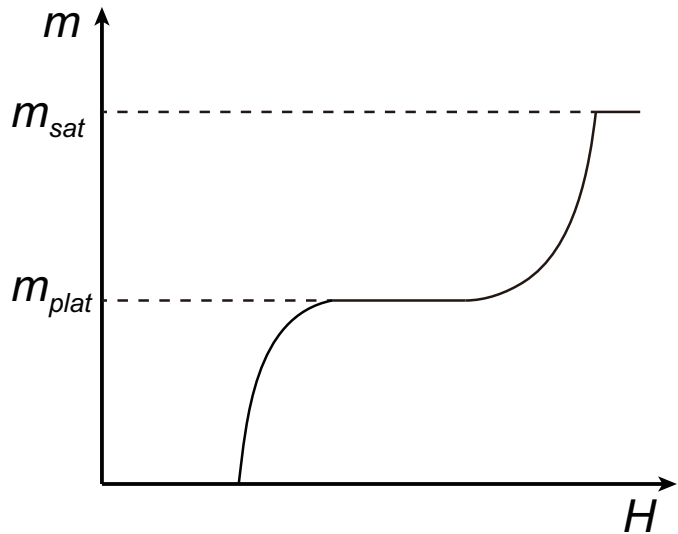}
\caption{
Schematic picture of a magnetization curve (magnetization per site $m$ as a function of applied magnetic field $H$) featuring a plateau region at $m=m_{plat}$. The saturation value of $m$ is 
denoted as $m_{sat}$.  
(adapted from Kim and Tanaka\cite{Kim})
}
\label{fig:mhcurve4}
\end{center}
\end{figure}

This can be understood in the following way (though the plateau can occur for 
general spatial dimensions, we continue to discuss the 2d case). 
Since the spin component parallel to 
the magnetic field (assume this to be in the z-direction), i.e. the magnetization, 
is already optimized by the field, it is not free to 
participate in the low energy spin fluctuations (in other words, the fluctuation in the field direction is energetically costly). 
The low energy effective theory  
therefore should consist of the dynamics that take place within the xy-plane, which can be 
represented by an angular field $\phi$, as in the superfluid action discussed above. 
The action\cite{Tanaka09} in fact turns out to have the same form as eq.(\ref{quantum XY action}), but with $\rho$ substituted by $\frac{S-m}{a^2}$, where $S$ is the spin quantum number, $m$ the magnetization per site, and $a$ the lattice constant. 
Therefore, in analogy with eq.(\ref{vortex phase factor SF}) a round excursion of a vortex 
enclosing an area $A$ which contains $N_{\rm sites}\in{\bf Z}$ sites will contribute a phase factor of 
\begin{equation}
e^{-{\cal S}_{\rm top}}=e^{-i2\pi \frac{S-m}{a^2} A}=e^{-i2\pi (S-m) N_{\rm sites}}.
\label{vortex phase factor plateau}
\end{equation} 
As before, for generic values of $S-m$ these Berry phase factors will suppress vortex condensation. Condensation may occur, if energetically favorable, 
when $S-m\in{\bf Z}$, which will disorder the ground state and create a finite energy gap 
between the ground state and the  excited states. This gapful state is none other than 
the magnetization plateau.

\subsection{Haldane gap}
Perhaps the most well known application of the interference effect of Berry phase or topological terms to condensed matter systems is the pioneering work 
by Haldane\cite{Haldane} on antiferromagnetic Heisenberg spin chains described by the 
Hamiltonian 
\begin{equation}
{\cal H}=J\sum_{j}\bol{S}_{i}\cdot\bol{S}_{j+1},
\end{equation}
where $J>0$. The order parameter for the problem is chosen to be 
the unit vector $\bol{n}_{j}$ 
which specifies the orientation of the staggered spin moment, 
i.e. $\bol{S}_{j}\approx S(-1)^j \bol{n}_{j}$. 

\begin{figure}[ht]
\begin{center}
\includegraphics[width=3.5in]{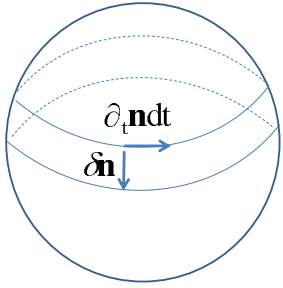}
\caption{Trajectories corresponding to two  
slightly varying (imaginary) time evolutions of the unit vector 
$\bol{n}(t)$. The difference in the solid angle that is traced out amounts to 
$\delta\omega=\int dt \bol{n}\cdot\delta\bol{n}\times\partial_t \bol{n}$.}
\label{fig:spherevariation}
\end{center}
\end{figure}

The kinetic term of the effective action should describe the tendency for 
nearby $\bol{n}_{j}$s to align, and is found to take the form 
\begin{equation}
{\cal S}_{\rm kin}=\int d\tau dx \frac{1}{2g}\left[(\partial_{\tau}\bol{n})^{2}+
(\partial_{x}\bol{n})^{2}
\right]
\end{equation}
in the continuum approximation, where $g$ is a coupling constant which depends on the details 
of the starting lattice Hamiltonian (we have again set an additional coefficient with the dimension of  velocity to unity for simplicity). 
In order to derive the topological term of the effective action, we 
begin by observing that the spin Berry phase term $iS\omega[\bol{n}(\tau)]$ for a single spin moment is odd with respect to spin inversion, i.e. 
$iS\omega[-\bol{n}(\tau)]=iS(4\pi-\omega[\bol{n}(\tau)])\equiv -iS\omega[\bol{n}(\tau)]$.  
The first equality follows by noting that the solid angle is an {\it oriented} 
surface area, whereas the second equality points to the fact that since this term always appears in the form of a phase factor, the portion $i4\pi S$ is irrelevant (due to $e^{-i4\pi S}\equiv 1$). Thus the contributions from the Berry phase term for 
each spin moment in the antiferromagnet add up into
\begin{eqnarray}
{\cal S}_{\rm top}&=&i\sum_{j}S(-1)^j \omega[\bol{n}_{j}(\tau)]\nonumber\\ 
&\approx& \frac{S}{2}\int dx \partial_{x}\omega[\bol{n}(\tau, x)].
\label{Haldane map}
\end{eqnarray}
On taking the continuum limit in the second line, we have converted differences into 
derivatives, and further used the fact that the derivatives are contributed by every 
other link on the chain (resulting in the factor of 1/2). 
To proceed, we seek the help of Fig.\ref{fig:spherevariation}, and obtain the final form of the 
topological term (which is often referred to in the literature of a $\theta$ term),
\begin{eqnarray}
{\cal S}_{\rm top}&=&i\frac{S}{2}\int d\tau dx \bol{n}\cdot\partial_{\tau}\bol{n}
\times\partial_{x}\bol{n}\nonumber\\
&\equiv& i\theta Q_{\tau x},
\end{eqnarray}
where $\theta\equiv2\pi S$, and 
\begin{equation}
Q_{\tau x}\equiv\frac{1}{4\pi}\int d\tau dx\bol{n}\cdot\partial_{\tau}\bol{n}\times
\partial_{x}\bol{n} \in{\bf Z},
\end{equation}
is an integer valued winding number which counts the number of times $\bol{n}(\tau, x)$ wraps around the sphere as one probes through the entire (Euclidean) space-time. The partition 
function therefore reads
\begin{eqnarray}
{\cal Z}&=&\int {\cal D}\bol{n}(\tau, x)e^{-{\cal S}[\bol{n}(\tau, x)]}\nonumber\\
&=&\sum_{Q_{\tau x}\in{\bf Z}}e^{-i2\pi S Q_{\tau x}}\int_{Q_{\tau x}} {\cal D}\bol{n}(\tau, x)
e^{-{\cal S}_{\rm kin}}.
\label{Haldane partition function}
\end{eqnarray}
In the second expression we have sorted the configurations entering the path integral 
according to the value of its winding number $Q_{\tau x}$. 
Eq.(\ref{Haldane partition function}) suggests that the system behaves in a qualitatively 
different way for integer $S$ (for which $e^{-i2\pi S Q_{\tau x}}=1$) 
and half odd integer $S$ (where $e^{-i2\pi S Q_{\tau x}}=(-1)^{Q_{\tau x}}$). 
For the latter case, the sign-alternating factors tend to lead to a destructive 
interference between configurations with nonzero $Q_{\tau x}$ (such space-time 
configurations are often called instantons). Since instantons will apparently 
cause a strong disruption to the antiferromagnet order, we thus  
expect that antiferromagnetic spin chains 
with half odd integer spin, for which case the instanton events are suppressed, 
should exhibit a stronger degree of spin ordering than those with integer spins, 
where instantons are {\it not} suppressed. It is by now well established that the former 
are indeed critically ordered (i.e. exhibit a power-law decaying spin-spin correlation), while 
the latter are strongly disordered (with exponentially decaying correlations).

\section{Quantum dynamics of topological objects}

An important aspect of topological terms is that they give rise to 
unconventional dynamics: since they often describe some variant 
of the Aharanov-Bohm effect, it is natural to expect that the 
behavior of the degree of freedom in motion under the influence of 
the fictitious ``magnetic field'' will 
differ from that of a free particle. 
This, for instance, lies at the heart of the anomalous Hall effect 
which is observed in frustrated ferromagnets. 

\subsection{Vortex motion in a superfluid/superconductor}
Here we discuss this aspect of topological 
terms in the simple context of a vortex in motion 
within a superconducting or superfluid thin film. (Though we focus 
again on the 2d case, generalization to 3d is straightforward.)  
Recall from our previous discussions that the topological term for this problem
reads (being interested here in the equation of motion, we are switching to the 
real time formalism) 
\begin{equation}
{\cal S}_{\rm top}=\int dtd\bol{r}\rho\partial_{t}\phi. 
\end{equation}
We assume for simplicity that the time dependence of the 
phase $\phi$ comes solely from the change in the 
vortex position $\bol{R}(t)$, 
i.e. $\phi=\phi(\bol{r}-\bol{R}(t))$. 
Noticing that $\frac{\partial\phi}{\partial\bol{R}}=-\nabla\phi$, 
 we can rewrite the topological term as 
\begin{equation}
{\cal S}_{\rm top}\equiv -q_{\rm v}\int dt \bol{v}\cdot \bol{A}(\bol{R}(t)),
\label{minimal coupling}
\end{equation}
where $\bol{v}\equiv \frac{d \bol{R}}{dt}$ is the vortex velocity, 
$q_{\rm v}$ is the vorticity, and 
\begin{equation}
\bol{A}(\bol{R}(t))\equiv\frac{1}{q_{\rm v}}\int d\bol{r}\nabla\phi(\bol{r}-\bol{R}(t)).
\end{equation}
Eq.(\ref{minimal coupling}) has precisely the same form as an action 
describing the coupling between a point charge of strength $q_{\rm v}$ 
moving with a velocity $\bol{v}$, with  
a magnetic field which is represented by the vector potential $\bol{A}(\bol{R})$. 
We therefore deduce \cite{Ao,Fisher Magnus} 
that the vortex suffers the pseudo-Lorentz force 
\begin{eqnarray}
\bol{F}&=&q_{\rm v}\bol{v}\times\left(\nabla_{\bol{R}}\times\bol{A}(\bol{R})
\right)
\nonumber\\
&=&q_{\rm v}\bol{v}\times2\pi \rho{\hat z}.
\end{eqnarray}
(An equivalent but more formal procedure, which will be used shortly, would be to resort to the 
principle of virtual work and calculate $\bol{F}\equiv -\frac{\delta {\cal S}_{\rm top}}{\delta \bol{R}}$.)
This is in full agreement with the physical picture arising from the 
AB effect which the 
Berry phase factor of eq.(\ref{vortex phase factor SF}) describes; the 
fictitious magnetic field is very much real to the vortex.  
The pseudo-Lorentz force plays an important role in the dynamics 
of vortices. For example it has been argued that it influences 
the Hall conductivity of the cuprate high temperature superconductors in the mixed state\cite{Feigelman}, where 
the role played by the bound states at the vortex core must also be incorporated.  

\subsection{Motion of skyrmions in 2d}
A very similar force also arises in the dynamics of {\it skyrmions}  
in 2d ferromagnets\cite{Stone}, which are topological objects that are recently 
receiving considerable attention due to their potential value toward 
applications.  
\begin{figure}[ht]
\begin{center}
\includegraphics[width=3.5in]{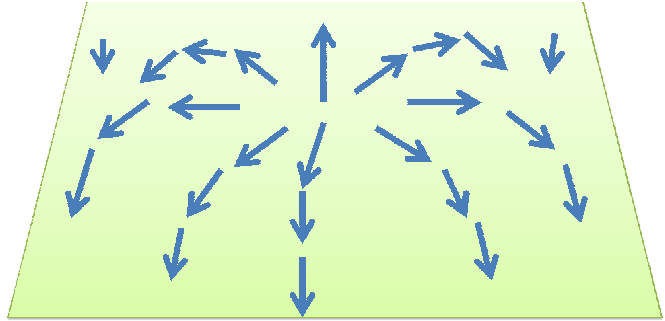}
\caption{Schematic illustration of a skyrmion configuration in a ferromagnet. Skyrmions are 
spin configurations for which the winding number $q_{\rm s}=\frac{1}{4\pi}\int dxdy \bol{n}\cdot\partial_{x}\bol{n}\times\partial_{y}\bol{n}\in{\bf Z}$, which counts the number of times the unit vector $\bol{n}(x, y)$ (indicated by the arrows) wraps around the unit sphere, is nonzero. In the configuration depicted, $q_{\rm s}=1$.} 
\label{fig:skyrmion}
\end{center}
\end{figure}
Here the topological term takes the form (in real time)
${\cal S}_{\rm top}=\int dtd^{2}\bol{r}\rho S\omega[\bol{n}(t, \bol{r})]$, where 
$\rho$ is the density of spins (whose spin quantum number is $S$ and its direction 
is represented by the unit vector $\bol{n}$) and $S\omega$ is the Berry phase term 
associated with a spin residing at position $\bol{r}$. 
The skyrmion is a configuration for which the winding number associated with the 
snapshot configuration (i.e. instantaneous configuration)
\begin{equation}
q_{\rm s}=\frac{1}{4\pi}\int d^2 \bol{r} \bol{n}\cdot\partial_{x}\bol{n}\times\partial_y \bol{n}
\in \bf{Z}
\label{skyrmion winding number}
\end{equation}
is a nonzero integer. 
(Notice that mathematically this is the same winding number (though associated 
with a different base manifold) 
which appeared in the previous section when we discussed the Haldane gap of antiferromagnetic spin chains.) A typical example of a skyrmion with $q_{\rm s}=1$ is 
depicted in Fig. 6. 
Since this integer valued number 
cannot change continuously, a skyrmion 
is stable unless processes involving singular configurations (which usually are energetically 
too costly to be relevant) are allowed. 
As in the vortex dynamics problem, we assume that the time dependence of the 
field takes the form $\bol{n}=\bol{n}(\bol{r}-\bol{R}(t))$, where $\bol{R}(t)$ now stands for 
the center of the skyrmion (or more generally a collective coordinate of the skyrmion).  
Using the formula for the variation of the surface angle (see Fig. 5)  
$\delta\omega=
\int dt \bol{n}\cdot\delta\bol{n}\times\partial_t \bol{n}$,  
we readily find that 
\begin{equation}
\bol{F}=-\frac{\delta {\cal S}_{\rm top}}{\delta \bol{R}}
=4\pi S q_{\rm s}\rho\bol{v}\times{\hat z}.
\label{force on skyrmion}
\end{equation}
Thus a skyrmion in motion behaves like a charged particle in the presence of 
a ``magnetic field'' $\propto \rho{\hat z}$. 

The construction of eq.(\ref{force on skyrmion}) generalizes easily to 
3d. Although the physical context where this situation may arise is as yet unclear, 
we outline its derivation, anticipating the future realization 
of a {\it 3d analog of the skyrmion}.  This also serves as an example on how 
topological phenomena occurring in different dimensions are intimately related 
through their mathematical structures. 
(Readers uninterested in technical details can skip the remainder of this section.) 
We construct our 3d system (using cartesian coordinates $(x, y, z)$) 
by imagining that for each $(y, z)$, there is a {\it 1d chain}  
extending in the z-direction. Furthermore, we assume that each of these 
1d chains accommodate an SU(2) Wess-Zumino-Witten model, 
which typically arises in studies of 
antiferromagnetic spin chains. The latter can equivalently be written as an O(4) 
nonlinear sigma model (i.e. a model featuring a unit-length vector field $\bol{n}$
with four components) with the topological term 
\begin{equation}
{\cal S}_{\rm top}=\frac{2\pi k}{Area(S^{3})}\int_{0}^{1}du\int dtdx \epsilon^{abcd}
{\tilde n}_{a}\partial_{u}{\tilde n}_{b}\partial_{t}{\tilde n}_{c}\partial_{x}
{\tilde n}_{d}, 
\label{S3 WZ term}
\end{equation}
where $k\in\bf{Z}$, $Area(S^{3})=2\pi^2$ is the surface area 
of the unit three-sphere, 
and $\tilde{\bol{n}}(u, t, x)$ is an extension of the field $\bol{n}(t, x)$ which 
satisfies 
\begin{eqnarray}
\tilde{\bol{n}}(u=0, t, x)&=&{}^{t}(0, 0, 0, 1),\nonumber\\  
\tilde{\bol{n}}(u=1, t, x)&=&\bol{n}(t, x)
\label{WZextension for S3}
\end{eqnarray}
and interpolates smoothly as a function of the auxiliary variable $u$ between 
these two limits. That this is a natural generalization of the spin Berry phase term 
becomes apparent when we rewrite the latter in an analogous manner, 
${\cal S}_{\rm BP}=\frac{2\pi k}{Area(S^{2})}\int_{0}^{1}du\int dt \tilde{\bol{n}}\cdot\partial_{u}\tilde{\bol{n}}
\times\partial_{t}\tilde{\bol{n}}$, where $k=2S$ and $Area(S^{2})=4\pi$. The way to 
extend the three-component unit-length vector $\bol{n}$ to $\tilde{\bol{n}}$ should 
be obvious if one follows the example of eq.(\ref{WZextension for S3}). 
The variation of eq.(\ref{S3 WZ term}) with respect to the change 
$\bol{n}\rightarrow\bol{n}+\delta\bol{n}$ yields an expression similar to that for 
the spin Berry phase term, and is given by 
\begin{equation}
\delta{\cal S}_{\rm top}
=\frac{2\pi k}{Area(S^{3})}\int dtdx
\epsilon^{abcd}n_{a}\delta n_{b}\partial_{t}n_{c}\partial_{x}n_{d}.
\end{equation}
With this preparation we come back to the coupled-chains situation 
for which the topological term from each chain add up into the form 
\begin{equation}
{\cal S}_{\rm top}^{\rm 3d}\equiv\int dydz \rho{\cal S}_{\rm top}[\bol{n}(t, x)]\vert_{y,z},
\end{equation} 
($\rho$ is the ``areal density'' (within the yz-plane) of these 1d extended objects),  
and consider a configuration in 3d ($xyz$) space 
for which the following winding number is a nonzero integer.  
\begin{equation}
q\equiv\frac{1}{Area(S^{3})}\int dxdydz\epsilon^{abcd}
n_{a}\partial_{x}n_{b}\partial_{y}n_{c}\partial_{z}n_{d}\in\bf{Z}.
\end{equation}
Introducing again a collective coordinate $\bol{R}$ which represents the 
position of this solitonic object (a 3d generalization of the 2d skyrmion), 
we arrive once more at the now familiar form 
of force, 
\begin{equation}
\bol{F}=-\frac{\delta {\cal S}_{\rm top}^{\rm 3d}}{\delta \bol{R}}
=2\pi qk\rho\bol{v}\times{\hat z}.
\end{equation}

\section{Surface states}
Some of the topological terms that we encounter can be expressed as a total divergence. 
Under periodic boundary conditions this property often allows us to associate 
the term with a topological winding number; this was the case for the topological term 
(usually referred to as a $\theta$ term) which appeared in the Haldane gap problem. 
However, actual physical systems have a surface, in which case a total divergence term 
will contribute a surface action. The appearance of surface effects which reflect the 
topology of the bulk (the bulk-surface correspondence) is another characteristic 
feature of systems which are governed by topological terms.    

\subsection{Haldane gap}
Let us revisit eq.(\ref{Haldane map}), the topological term 
for a spin $S$ antiferromagnetic spin chain. 
We saw that under a periodic boundary condition,  
this lead to a $\theta$ term which discriminates between the behavior of integer and half-odd integer $S$. Let us now consider instead a spin chain with {\it open ends}. 
The topological term apparently reads, upon carrying out the 
spatial integration, 
\begin{equation}
{\cal S}_{\rm top}=i\frac{S}{2}\omega[\bol{n}(\tau, x_{\rm R})]
-i\frac{S}{2}\omega[\bol{n}(\tau, x_{\rm L})],
\end{equation}
where $x_{\rm R}$ and $x_{\rm L}$ are the two edges of the chain. 
One recognizes that the two terms appearing in the right hand side are precisely 
the Berry phase terms for two spin moments, each residing at $x=x_{\rm R}$ 
and $x=x_{\rm L}$,  and carrying a spin quantum number of $S/2$. 
This suggests that {\it fractional spin} degrees of freedom  
are induced at the ends of a finite-length spin chain\cite{Ng}. 
The emergence of $S=1/2$ edge spins in $S=1$ antiferromagnets has been 
verified experimentally as well as numerically. It should be noted that 
this simple argument is only valid for integer $S$, where there is a spectral gap 
and an associated finite correlation length (the bulk of the system is unaffected by the 
presence of the boundary). For half odd integer $S$, with an infinite correlation 
length, the chain end effects are more subtle and require a detailed treatment. 
  
\subsection{Topological insulator}
As there are by now a considerable number of accessible reviews on topological insulators\cite{Kane}, 
we will not go into their details, and merely remark that here also, several features similar to 
the edge state physics of Haldane gap systems appear. One example is the 
electromagnetic Hall response at the surface. The 
action governing the gauge response of a 3d topological insulator with a time reversal symmetry contains the unconventional term (which is also called a $\theta$ term, 
or an axion term depending on how one interprets the action) 
\begin{equation}
{\cal S}_{\rm top}=i\int d\tau d\bol{r}\theta\frac{e^{2}}{2\pi h}\bol{E}\cdot\bol{B}.\end{equation}
(Readers may wonder why this term becomes purely imaginary when turning to the 
imaginary time framework. This can be understood by recalling that  $A_{0}$, 
the time component of the electromagnetic gauge field (i.e. the scalar potential) 
should transform like $\frac{\partial}{\partial t}$, and thus $A_{0}\rightarrow iA_{0}$ and accordingly $\bol{E}\rightarrow i\bol{E}$ (while 
$\bol{B}\rightarrow\bol{B}$) upon  
switching to Euclidean space-time.) 
It is well known that this term, like the $\theta$ term of the spin chain problem, 
is a total divergence, if $\theta$ is constant. More precisely, it is the divergence of 
the abelian Chern-Simons term which famously describes a quantum Hall response. 
From the requirement of time reversal symmetry, it can be shown that $\theta$ can only 
be (modulo $2\pi$) either $0$ (conventional insulator) or $\pi$ (topological insulator). 
This is analogous to the fact that the coefficient $\theta$ in the spin chain problem 
is also restricted to these two values due to spatial inversion symmetry (unless a bond 
alternating exchange interaction is introduced). 
Since the vacuum can be regarded as a conventional insulator, it follows that 
there is a jump in this value by the amount $\Delta\theta=\pi$ 
at the surface of a topological insulator. Thus there arises a surface Chern-Simons term 
which yields a half-integer quantized Hall effect $\sigma_{xy}=\e^{2}/2h$.The observation 
of this effect requires that the surface state be gapped by a suitable perturbation,  
such as placing a magnetic thin film on the sample surface\cite{Kane}. 

Finally we note that a topological field theory describing the physics 
of the {\it matter field} of topological insulators has also been proposed\cite{BF}.
   
\section{Topological structure of wavefunctions}

In the previous section, we have seen that when the topological term is a total divergence, 
it can induce an edge state at the spatial surface of the system. As was noted by Xu and 
Senthil\cite{Xu}, a similar surface contribution will arise at a {\it temporal surface} 
if the situation at hand requires that we have an open boundary in the (imaginary) time direction. Recall in this regard, that the path integral representation of a ground state 
wavefunction which we mentioned back in section 2 meets this condition: since 
we need to specify the initial and final boundary data of the field, the temporal boundary condition is by construction not periodic. 

The way in which temporal surface terms influence the ground state wavefunction (or wavefunctional) 
has interesting applications\cite{Xu} to the study of {\it symmetry protected topological   states} (SPT states)\cite{Chen}, which are the conceptual generalization of topological insulators (the basis of 
which rests on the framework of noninteracting electrons) to interacting systems. 
The concept of SPT states need not be restricted to electronic systems but can also apply 
as well for instance to bosonic and magnetic systems. 
As our final application of topological terms, we will 
briefly revisit the 1d version of the magnetization plateau problem of section 3.3 from this perspective, highlighting results from our recent work\cite{Takayoshi}. 
For details we refer the reader to the original reference. 

 We will hereafter concentrate exclusively 
on the magnetization plateau state, which according to our earlier discussion imposes 
on our system the condition $S-m\in{\bf Z}$. 
Let us also recall that basically, the low energy physics 
only involves the spin fluctuation within the plane perpendicular to the applied 
magnetic field. This means that out of the total spin moment $S$ 
residing at each site of our 1d system, a portion $m$ is segregated off into a 
higher energy sector by the external field. It therefore follows that  
in the low energy sector, we are left to deal 
with an effective subsystem consisting of a planar antiferromagnetic chain 
with spin quantum number $S-m (\in {\bf Z})$ in the {\it absence} of a magnetic field 
(since the planar spin component is unaffected by the Zeeman coupling). With this substantial simplification in hand, we can make contact with the effective action 
for the Haldane gap system of section 3.4, 
where we are to make the following two modifications: 
(1) the spin quantum number $S$ should be replaced by $S-m$, and 
(2)the planar limit must be taken. Carrying out the second modification properly 
is a subtle procedure which has been detailed elsewhere\cite{Takayoshi,Kim}, and 
we will go directly to the result for the imaginary-time effective action obtained by the strategy just 
outlined:
\begin{eqnarray}
{\cal S}&=&{\cal S}_{\rm kin}+{\cal S}_{\rm top}\nonumber\\
{\cal S}_{\rm kin}&=&\int d\tau dx \frac{1}{2g}\left(\partial_{\mu}\phi\right)^{2}
\nonumber\\
{\cal S}_{\rm top}&=&i(S-m)\int d\tau dx \left(
\partial_{\tau}a_{x}-\partial_{x}a_{\tau}\right)
\nonumber\\
a_{\mu}&\equiv&\frac{1}{2}\partial_{\mu}\phi  {\mbox{\hspace{2mm}}}(\mu=\tau, x). 
\end{eqnarray} 
Notice that this action is a total divergence and will, in the presence of an open boundary,  generate surface contributions. 
For the sake of simplicity we will take the strong coupling limit $g\rightarrow\infty$, so that the action is dominated by the topological term.  
Using the prescriptions of section 2, we find that the ground state wavefunctional takes the form (we employ a periodic boundary condition in the {\it spatial} direction)
\begin{equation}
\Psi[\phi(x)]\propto
\int_{\phi_{\rm i}\rightarrow\phi(x)}{\cal D}\phi(\tau, x)
e^{-i(S-m)\int dx a_{x}(\tau_{\rm f})-a_{x}(\tau_{\rm i})}.
\end{equation}
The action appearing in the exponent are the temporal surface terms mentioned at the 
outset of this section. 
Since the initial data is not relevant for our purpose, we need 
only concern ourselves with the final data, which results in 
\begin{equation}
\Psi[\phi(x)]\propto e^{-i\pi(S-m)W},
\label{wavefunctional}
\end{equation} 
where $W\equiv\frac{1}{2\pi}\int dx\partial_{x}\phi\in{\bf Z}$ 
is a winding number recording the number of $2\pi$ rotations that the 
angular field $\phi$ executes along the x-axis in the snapshot configuration $\phi(x)$. 
The dependence of eq.(\ref{wavefunctional}) on $W$ differs dramatically depending on 
whether $S-m$ is odd ($\Psi\propto(-1)^W$) or even ($\Psi\propto 1$). 
Since this structure is only dependent on the topological winding number $W$, 
it is expected to persist when perturbations of moderate strength are applied  
to the system. 
The fact that there are no ways to adiabatically connect these two wavefunctionals 
suggests that they describe different phases. (Furthermore, given that 
the topological term plays no role in determining the global structure of $\Psi$ 
(i.e. the dependence on the topological sector labeled by $W$)  
for even $S-m$, we expect that the ground state belongs to a 
topologically trivial phase, while the $S-m$ odd system is expected to lie in a nontrivial phase). 
This expectation is confirmed numerically 
as well as by a rigorous analysis of a solvable model\cite{Takayoshi}. 

Finally let us turn to the symmetry aspects of the present problem. Due to the presence of a 
magnetic field, time reversal symmetry is explicitly broken. Meanwhile the system does respect the symmetry with respect to an inversion about the center 
of a link connecting adjacent spins. To see how this symmetry is affecting the ground state, 
we break it by switching on a staggered magnetic field.  This perturbation will induce a 
staggered magnetization $(-1)^{j}\delta m$ (where $j$ is the site index) along the chain. 
By repeating the evaluation of the ground state wavefunctional for this case, we find that 
it now modifies to 
\begin{equation}
\Psi[\phi(x)]\propto e^{-i\pi(S-m-\delta m)W}. 
\end{equation} 
We thus find that by sweeping the value of $\delta m$ by tuning the staggered magnetic field, 
we can smoothly interpolate between the two types of wavefunctionals 
which was not possible in the absence of the staggered field (i.e. as long as the link-center inversion symmetry was respected). 
By employing a duality transformation technique, we can further show\cite{Takayoshi,Kim}  
that it is possible to carry out this interpolation without encountering 
a gap-closing point (i.e. a quantum phase transition), which indicates that the two 
groundstates now belong to the same phase. 
We thus conclude that the odd $S-m$ plateau is: (1) a topologically nontrivial 
state belonging to a phase distinct from an even $S-m$ plateau state (2) which is 
{\it protected} by link-centered inversion symmetry. This is similar to the topological 
insulator story: the 3d ${\bf Z}_{2}$ topological insulator, which belongs to a phase 
to be distinguished from a conventional insulator, owes its integrity and robustness 
to the time reversal  symmetry. Once that symmetry is broken, the distinction between 
the topological and conventional insulators is lost.  

We thus see that the recent development centering around topological insulators and superconductors provides us with new insights for exploring topological effects (especially 
on their robustness) in a broader range of condensed matter systems, 
including those which have been known for several decades. 
Most of the physical examples of SPT states 
identified so far are 1d systems. The identification of higher dimensional SPT states, 
along with their complete characterization and classification is an important problem 
left for the future.

\section{Acknowledgments}
Portions of this article are based on 
a series of lectures which AT presented at the 
Graduate School of 
Advanced Science of Matter at Hiroshima University. 
Thanks goes to K.-I. Imura for the warm hospitality during that occasion. 
The work of AT was supported in part by KAKENHI no.(C) 23540461.

\end{document}